\documentclass{article}
\usepackage{spconf,amsmath,graphicx,hyperref,booktabs,afterpage,etoolbox}
\usepackage{orcidlink}

\makeatletter
\renewcommand\section{\@startsection{section}{1}{\z@}%
  {-2.0ex \@plus -.5ex \@minus -.2ex}%
  {1.0ex \@plus .1ex}%
  {\normalfont\bfseries}}
\renewcommand\subsection{\@startsection{subsection}{2}{\z@}%
  {-1.7ex \@plus -.4ex \@minus -.2ex}%
  {0.8ex \@plus .1ex}%
  {\normalfont\bfseries}}
\renewcommand{\name}[1]{\gdef\@name{{\em #1}\rule[-1.2em]{0pt}{1em}}}
\patchcmd{\@maketitle}{\par\vskip 1.5em}{\par\vskip 0.7em}{}{}
\makeatother

\title{OpenPA: Forced-Alignment-Free Multi-Granular Pronunciation Assessment with Mixed-Transcript Adaptation}
\name{Fu-An Chao\,\orcidlink{0000-0003-2172-8001},
Bi-Cheng Yan\,\orcidlink{0009-0002-1987-2418},
Berlin Chen\,\orcidlink{0000-0003-0693-8932}}
\address{National Taiwan Normal University, Taipei, Taiwan \\
{\small\texttt{\{fuanchao, bicheng, berlin\}@ntnu.edu.tw}}
}
\begin{document}
%\ninept
%
\maketitle
\begin{abstract}
Recent advances in automatic pronunciation assessment (APA) have driven a shift from scripted to free-form assessment.
However, existing methods often rely on multi-stage pipelines with forced alignment and separate feature extractors, limiting their practicality across response settings.
In light of this, we propose \textbf{OpenPA}, a novel APA framework that leverages Whisper's cross-attention mechanism to jointly predict word- and utterance-level pronunciation scores without forced alignment. 
Through mixed-transcript adaptation and adaptive substitution weighting, OpenPA accommodates heterogeneous transcripts across closed- and open-response settings within a single model.
Trained solely on speechocean762, OpenPA achieves the strongest overall performance among evaluated systems under in-domain closed and simulated open-response conditions, while maintaining its performance zero-shot on MultiPA.
An additional cross-task evaluation on oral-cancer speech suggests its potential applicability beyond non-native speech.
\end{abstract}
\section{Introduction}
\label{sec:intro}

As a pivotal component of computer-assisted pronunciation training (CAPT),
automatic pronunciation assessment (APA) aims to provide learners with fine-grained and
multifaceted pronunciation feedback~\cite{chao2025hmamba}. Conventional APA systems have primarily focused on the assessment of scripted
read-aloud speech~\cite{gong2022gopt, chao20223m, do2023hipama}, where learners follow a predefined reference text in a closed-response setting. 
More recently, research has increasingly shifted towards unscripted speech~\cite{chen2024multipa, chen2025read, yan2025hippo},
allowing learners to respond to questions freely without predefined text prompts. 
Although this open-response setting better reflects communicative speaking, it poses new
challenges beyond those encountered in read-aloud assessment.

% Queue Fig. 1 as a top-of-column float. Since it is encountered after text has
% begun in the left column, LaTeX places it at the top of the right column while
% allowing the Introduction to continue beneath it.
\begin{figure}[t]
  \centering
  \setlength{\abovecaptionskip}{0pt}
  \setlength{\belowcaptionskip}{0pt}
  \includegraphics[width=0.88\columnwidth]{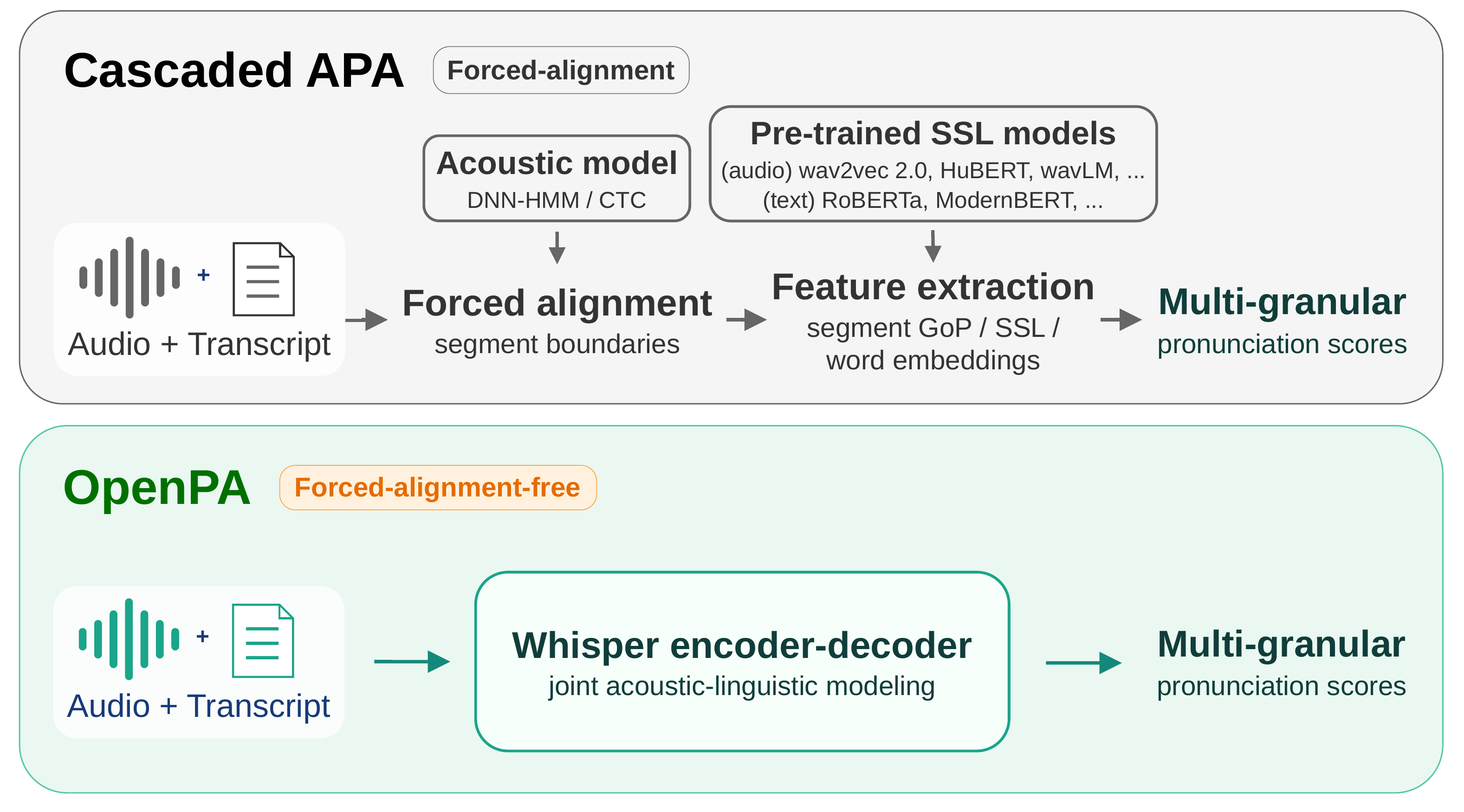}
  \caption{Conventional alignment-based APA vs. OpenPA.}
  \label{fig:others-vs-openpa}
\end{figure}

% Defer the result tables until page 3 so they remain together at its top.
\afterpage{\afterpage{\begin{table*}[t]
  \centering
  \setlength{\abovecaptionskip}{0pt}
  \setlength{\belowcaptionskip}{2pt}
  \caption{In-domain PCC on speechocean762. ASR and ground-truth transcripts
  represent simulated open- and closed-response conditions, respectively.
  Open-response word PCC follows M+S/M/S
  (matched+substituted/matched/substituted). Overall Avg. PCC averages
  10 reported PCCs for ASR input and 6 for ground-truth input.}
  \label{tab:speechocean-results}
  \footnotesize
  \setlength{\tabcolsep}{2pt}
  \renewcommand{\arraystretch}{0.82}
  \begin{tabular*}{\textwidth}{@{\extracolsep{\fill}}llccccccc@{}}
    \toprule
    Transcript & System &
    \multicolumn{2}{c}{Word level} &
    \multicolumn{4}{c}{Utterance level} & Overall \\
    \cmidrule(lr){3-4}\cmidrule(lr){5-8}
    & & Accuracy↑ & Total↑ & Accuracy↑ & Fluency↑ & Prosodic↑ & Total↑ & Avg. PCC↑ \\
    \midrule
    & MultiPA$^{*}$~\cite{chen2024multipa} & 0.397\,/\,0.271\,/\,0.274 & 0.406\,/\,0.291\,/\,0.281
    & 0.699 & 0.778 & 0.772 & 0.718 & 0.489 \\
    medium.en {\scriptsize (WER:20.1\%↓)} & HiPPO$^{*}$~\cite{yan2025hippo} & 0.313\,/\,0.285\,/\,0.042 & 0.313\,/\,0.294\,/\,0.041
    & 0.720 & 0.792 & 0.785 & 0.741 & 0.433 \\
    & \textbf{OpenPA} & \textbf{0.506\,/\,0.451\,/\,0.295} & \textbf{0.516\,/\,0.456\,/\,0.308}
    & \textbf{0.758} & \textbf{0.801} & \textbf{0.791} & \textbf{0.764} & \textbf{0.565} \\
    \midrule
    & MultiPA$^{*}$~\cite{chen2024multipa} & 0.405\,/\,0.324\,/\,0.308 & 0.416\,/\,0.342\,/\,0.318
    & 0.705 & 0.781 & 0.779 & 0.725 & 0.510 \\
    large-v3 {\scriptsize (WER:16.3\%↓)} & HiPPO$^{*}$~\cite{yan2025hippo} & 0.331\,/\,0.329\,/\,0.052 & 0.330\,/\,0.336\,/\,0.050
    & 0.724 & 0.793 & 0.786 & 0.745 & 0.448 \\
    & \textbf{OpenPA} & \textbf{0.526\,/\,0.504\,/\,0.335} & \textbf{0.534\,/\,0.512\,/\,0.347}
    & \textbf{0.758} & \textbf{0.803} & \textbf{0.794} & \textbf{0.765}
    & \textbf{0.588} \\
    \midrule
    & MultiPA$^{*}$~\cite{chen2024multipa} & 0.512 & 0.524 & 0.731 & 0.799 & 0.795 & 0.750 & 0.685 \\
    ground-truth & HiPPO$^{*}$~\cite{yan2025hippo} & 0.630 & 0.640 & 0.781 & \textbf{0.829}
    & \textbf{0.825} & 0.803 & 0.752 \\
    & \textbf{OpenPA} & \textbf{0.714} & \textbf{0.719} & \textbf{0.800}
    & 0.826 & 0.817 & \textbf{0.804} & \textbf{0.780} \\
    \bottomrule
  \end{tabular*}

\vspace{0.15em}

\begin{minipage}[t]{0.48\textwidth}
  \centering
  \caption{Zero-shot cross-dataset PCC on MultiPA using medium.en transcripts; dashes denote unreported results.}
  \label{tab:multipa-results}
  \footnotesize
  \setlength{\tabcolsep}{2pt}
  \renewcommand{\arraystretch}{1.22}
  \begin{tabular*}{\columnwidth}{@{\extracolsep{\fill}}lcccc@{}}
    \toprule
    System & Word↑ & Accuracy↑ & Fluency↑ & Prosodic↑ \\
    \midrule
    MultiPA$^{*}$~\cite{chen2024multipa} & 0.346 & 0.583 & 0.649 & 0.473 \\
    TextPA(gpt-4o-mini)~\cite{chen2025read} & -- & 0.728 & 0.650 & 0.243 \\
    TextPA(gpt-4o-mini)+MultiPA~\cite{chen2025read} & -- & 0.769 & 0.784 & -- \\
    \textbf{OpenPA} & \textbf{0.452} & \textbf{0.783}
    & \textbf{0.810} & \textbf{0.644} \\
    \bottomrule
  \end{tabular*}
\end{minipage}
\hfill
\begin{minipage}[t]{0.48\textwidth}
  \centering
  \caption{Zero-shot cross-task speaker-level PCC on the PathBench YouTube oral-cancer benchmark.}
  \label{tab:pathbench-results}
  \footnotesize
  \setlength{\tabcolsep}{4pt}
  \renewcommand{\arraystretch}{0.93}
  \vspace{-6pt}
  \begin{tabular*}{0.48\columnwidth}[t]{@{\extracolsep{\fill}}lc@{}}
    \toprule
    \multicolumn{2}{c}{\textit{Reference-Free (Model)}} \\
    \cmidrule(lr){1-2}
    System & PCC↑ \\
    \midrule
    ASRIC~\cite{halpern2026pathbench} & 0.55 \\
    DArtP~\cite{halpern2026pathbench} & 0.78 \\
    Confidence~\cite{halpern2026pathbench} & 0.79 \\
    \textbf{OpenPA (Fluency)} & \textbf{0.82} \\
    \bottomrule
  \end{tabular*}
  \hfill
  \begin{tabular*}{0.48\columnwidth}[t]{@{\extracolsep{\fill}}lc@{}}
    \toprule
    \multicolumn{2}{c}{\textit{Reference-Text}} \\
    \cmidrule(lr){1-2}
    System & PCC↑ \\
    \midrule
    PER (SEM)~\cite{halpern2026pathbench} & 0.66 \\
    PER (Phone)~\cite{halpern2026pathbench} & 0.75 \\
    ArtP~\cite{talkar2025artp} & 0.78 \\
    \textbf{OpenPA (Fluency)} & \textbf{0.79} \\
    \bottomrule
  \end{tabular*}
\end{minipage}
\end{table*}}}

% Place the qualitative visualization at the top of page 4.
\afterpage{\afterpage{\afterpage{\afterpage{\begin{figure*}[t]
  \centering
  \setlength{\abovecaptionskip}{-20pt}
  \setlength{\belowcaptionskip}{13pt}
  \makebox[\textwidth][c]{%
    \resizebox{\textwidth}{1.80in}{\includegraphics{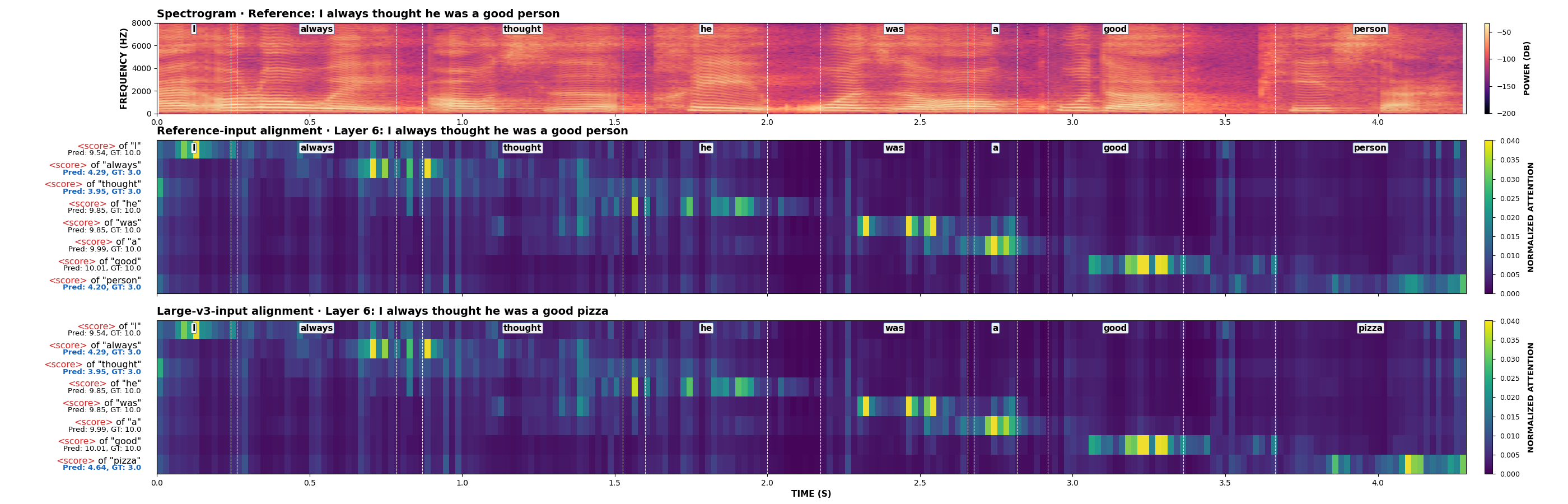}}%
  }
  \caption{Cross-attention under reference and large-v3 transcripts; predicted
  and ground-truth word scores appear on the left.\protect\footnotemark}
  \label{fig:qualitative-attention}
\end{figure*}}}}}

\vspace{0pt}
A central challenge in open-response APA is the scarcity of fine-grained
annotated training data. Although MultiPA~\cite{chen2024multipa} provides
word- and utterance-level annotations for real-world open responses, its 50 utterances
are better suited to pilot evaluation than model training. Therefore, existing approaches typically learn fine-grained
scoring from closed-response corpora such as speechocean762~\cite{zhang2021speechocean762} and evaluate
cross-dataset transfer on MultiPA.

% Queue the double-column architecture figure only after the first-page column
% break, so it remains at the top of page 2 without displacing Fig. 1.
\begin{figure*}[t]
  \centering
  \setlength{\abovecaptionskip}{-6pt}
  \setlength{\belowcaptionskip}{8pt}
  \includegraphics[width=0.9\textwidth]{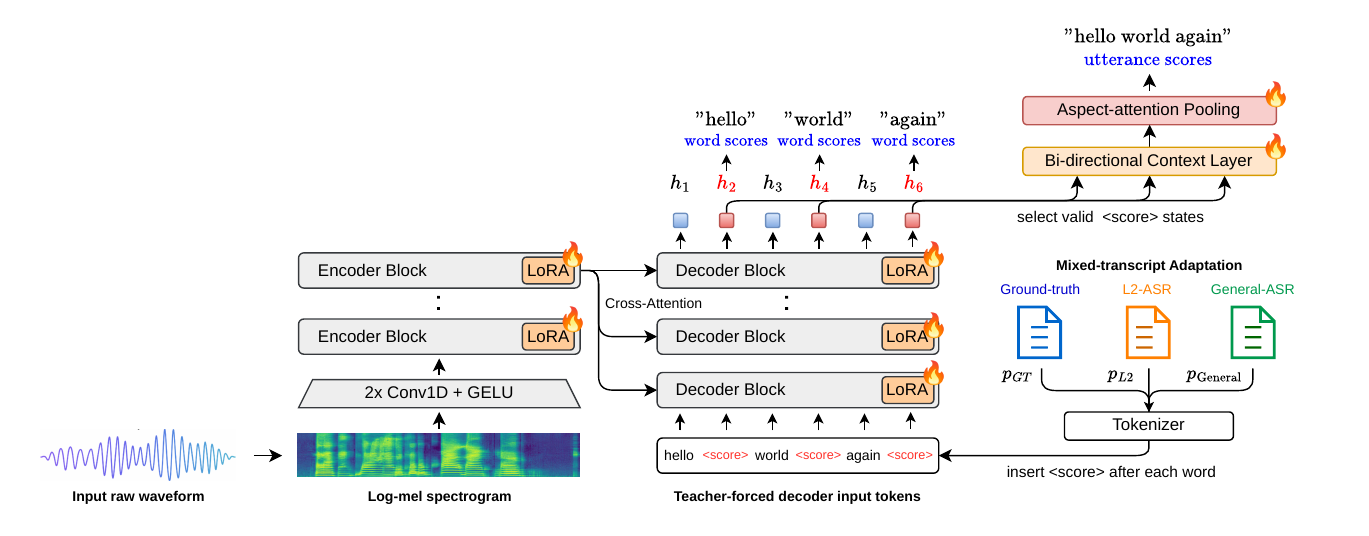}
  \caption{Overview of the proposed OpenPA framework.}
  \label{fig:openpa-architecture}
\end{figure*}

To assign fine-grained scores with
individual speech segments, prior efforts have predominantly relied on
forced alignment before extracting heterogeneous features with separate
scoring components~\cite{chen2024multipa, yan2025hippo}. Such cascaded pipelines may substantially
impede end-to-end optimization and deployment across different response
settings.

Building on these observations, we propose \textbf{OpenPA}, a unified, forced-alignment-free framework for
word- and utterance-level pronunciation assessment. As illustrated in
Fig.~\ref{fig:others-vs-openpa}, OpenPA replaces explicit forced alignment and separate
feature extraction with joint acoustic-linguistic modeling. Given evidence
that precise phoneme boundaries are not strictly required and may introduce alignment
errors~\cite{cao2024ctc}, OpenPA leverages Whisper's cross-attention~\cite{radford2023whisper} to associate
acoustic evidence with phonetic transcript tokens. Meanwhile, mixed-transcript adaptation and
adaptive substitution weighting improve robustness to ASR errors, supporting
both closed- and open-response assessment within a single model.

The main contributions of this paper are at least twofold: 
1) to our knowledge, OpenPA is the first forced-alignment-free framework 
for multi-granular pronunciation assessment at both the word and utterance levels, 
supporting either ground-truth or ASR-generated transcripts. 
2) we establish a consistent multi-granular open-response
benchmark and release its evaluation code~\footnote{\texttt{https://github.com/fuann/open-apa}}.
Experiments cover closed- and simulated open-response conditions on
speechocean762 and zero-shot cross-dataset evaluation on MultiPA.

\section{OpenPA}
\label{sec:format}

% TODO: Confirm that the final experimental command enables
% --utterance-context-layer, --utterance-pooling attention,
% --mixed-transcript-training, and --adaptive-substitution-loss. The current
% openpa-main/run_small.sh and run.sh defaults do not enable all four options.

\subsection{Forced-alignment-free pronunciation assessment}

As illustrated in Fig.~\ref{fig:openpa-architecture}, OpenPA is a
hierarchical pronunciation assessment model built upon Whisper. Given an
acoustic input $\mathbf{X}$ and its teacher-forced transcript
$\mathbf{T}=(w_1,\ldots,w_N)$, where each word $w_i$ may comprise multiple
BPE tokens, we introduce a special \texttt{<score>} token as a word-level
prediction representation, analogous to the \texttt{[CLS]} token in
BERT~\cite{devlin2019bert}. Specifically, we construct an augmented token
sequence $\widetilde{\mathbf{T}}$ by inserting \texttt{<score>} after the final
BPE token of each word.
Let $\mathbf{h}_t$ denote the decoder hidden state at token position $t$. For
the $i$-th word, its representation $\mathbf{r}_i$ is the hidden state at the
corresponding \texttt{<score>} position $p_i$. $\mathbf{r}_i$ is conditioned on the complete
sequence of acoustic encoder states $\mathbf{Z}=\operatorname{Enc}(\mathbf{X})$
through cross-attention and on the transcript prefix up to $w_i$ through causal
self-attention:

\begin{equation}
    \mathbf{r}_i = \mathbf{h}_{p_i}
      = \operatorname{Dec}_{p_i}
         (\widetilde{\mathbf{T}}_{\le p_i},\mathbf{Z}).
\end{equation}

Since \texttt{<score>} follows the word, this state summarizes
its lexical content, preceding linguistic context, and pronunciation-relevant
acoustic evidence. It can attend directly over the encoder sequence and
therefore does not require externally supplied word timestamps. Subsequently,
an aspect-specific regression head $g_k$ maps $\mathbf{r}_i$ to the predicted
score $\hat{s}_{i,k}$ for pronunciation aspect $k$ of word $w_i$:

\begin{equation}
    \hat{s}_{i,k} = g_k(\mathbf{r}_i).
  \label{eq:word-prediction}
\end{equation}

For utterance-level assessment, a lightweight bidirectional context layer
transforms the word representations $(\mathbf{r}_1,\ldots,\mathbf{r}_N)$ into
contextualized representations $(\mathbf{c}_1,\ldots,\mathbf{c}_N)$. Implemented
as a single Transformer encoder layer with multi-head self-attention and a
feed-forward network, it allows each word representation to incorporate
information from all words in the utterance. Inspired by HiPAMA~\cite{do2023hipama}, the contextualized representations
are then aggregated using aspect-specific attention pooling.
Specifically, it computes aspect-specific weights
$\alpha_{i,k}=\operatorname{softmax}_{i}(a_k(\mathbf{c}_i))$, 
where $a_k$ is the attention scorer for aspect $k$, and aggregates the
representations as $\mathbf{u}_k=\sum_{i=1}^{N}\alpha_{i,k}\mathbf{c}_i$ for
utterance-level scoring.
An aspect-specific regression head $f_k$ then maps $\mathbf{u}_k$ to the
corresponding utterance-level score:
\begin{equation}
    \hat{y}_k = f_k(\mathbf{u}_k).
  \label{eq:utterance-prediction}
\end{equation}
This hierarchy shares the Whisper model across both granularities, 
allowing utterance-level predictions to build on word-level evidence 
and capture cross-granularity dependencies.

\subsection{Mixed-transcript adaptation}

Open-response assessment is particularly challenging due to transcript uncertainty. Unlike closed response, it has no human-annotated transcript 
at inference time and instead relies on an ASR-generated transcript, 
making pronunciation scores susceptible to recognition errors. 
To alleviate this transcript mismatch, we perform mixed-transcript adaptation during training. 

\noindent\textbf{Transcript sampling:} 
For each utterance, we first sample the ground-truth, L2-adapted ASR, or general-purpose ASR transcript 
with probabilities $p_{GT}$, $p_{L2}$, and $p_{general}$, which are set to $(0.5,0.3,0.2)$, respectively. 
Ground-truth transcripts represent closed-response input, while the L2-adapted and
general-purpose hypotheses are generated by an off-the-shelf L2 ASR model
fine-tuned on the publicly-available S\&I corpus~\cite{knill2025sandi} and Whisper medium.en,
respectively (WER on the speechocean762 training set: 18.5\% and 21.6\%). 
The sampled transcript is teacher-forced into the decoder,
exposing OpenPA to both closed- and open-response conditions.

\noindent\textbf{Adaptive substitution weighting:}
However, adapting with ASR-generated transcripts requires transferring the available
human scores to the recognized words. To this end, we use Levenshtein alignment solely 
to construct training targets by mapping each ASR transcript to its human-annotated reference. 
Matched and substituted ASR words inherit their corresponding reference scores; insertions are masked, 
while deletions produce no decoder token and thus contribute no word-level loss. 

During optimization, a matched word receives unit weight. 
For an ASR substitution, however, the learner's intended word is intrinsically ambiguous to determine: 
the ASR output may reflect either a recognition error despite correct pronunciation 
or a genuine pronunciation deviation from the prompt.
The latter provides a strong pronunciation signal that should be emphasized during adaptation. 
Since the cause is unobserved, we use the reference word accuracy $a_i\in[0,10]$ as a proxy and assign
\begin{equation}
  \omega_i=\omega_{\min}+(1-\omega_{\min})(1-a_i/10).
  \label{eq:substitution-weight}
\end{equation}
Thus, with $\omega_{\min}=0.5$, substitutions associated with poorly pronounced reference words receive greater weight,
emphasizing likely pronunciation deviations while downweighting probable recognition errors.

\subsection{Joint optimization}

To preserve the quality of Whisper's pretrained decoder representations and
its original ASR capability, OpenPA retains the ASR generation objective while
jointly optimizing pronunciation assessment. We train OpenPA using LoRA adapters throughout the Whisper, together with the task-specific scoring modules. Let $\mathcal{L}_{\mathrm{ASR}}$
denote token-level cross-entropy. The word- and utterance-level losses,
$\mathcal{L}_{\mathrm{word}}$ and $\mathcal{L}_{\mathrm{utt}}$, are mean squared
errors, with $\mathcal{L}_{\mathrm{word}}$ incorporating the adaptive weights
in Eq.~\eqref{eq:substitution-weight}. The joint objective is
\begin{equation}
  \mathcal{L}=\lambda_{\mathrm{ASR}}\mathcal{L}_{\mathrm{ASR}}
  +\lambda_{\mathrm{word}}\mathcal{L}_{\mathrm{word}}
  +\lambda_{\mathrm{utt}}\mathcal{L}_{\mathrm{utt}}.
  \label{eq:joint-objective}
\end{equation}

\section{Experiments}
\label{sec:experiments}

\subsection{Experimental setup}

\noindent\textbf{Training:} We use speechocean762~\cite{zhang2021speechocean762},
which contains scripted English utterances from 250 non-native Mandarin
speakers with multi-granular pronunciation annotations. Following
\cite{chao2026probing}, OpenPA uses Whisper-medium as its backbone and is
fine-tuned for 5 epochs with rank-16 LoRA, a batch size of 64, and a learning
rate of $8\times10^{-4}$.

\noindent\textbf{Evaluation regimes:} We evaluate OpenPA under three regimes.
\textbf{In-domain:} The speechocean762 test set is evaluated under closed- and 
simulated open-response conditions using ground-truth and ASR-generated transcripts, respectively.
\textbf{Cross-dataset:} MultiPA's 50 real-world open responses~\cite{chen2024multipa} are used 
to assess zero-shot transfer, following its original Whisper medium.en transcript protocol.
\textbf{Cross-task:} An exploratory zero-shot task transfer is further evaluated on the PathBench~\cite{halpern2026pathbench} 
subset of the YouTube oral-cancer corpus~\cite{halpern2023oral}. 
The subset comprises 98 utterances from 21 speakers; utterance-level fluency predictions are averaged per speaker.
Neither out-of-domain dataset is used during training.
In the following, we refer MultiPA to the dataset; MultiPA$^{*}$ and HiPPO$^{*}$ are existing forced-alignment approaches reproduced by
author-provided pretrained models; other compared results are derived from the original papers.

\noindent\textbf{Word-level evaluation:} Because open-response word-level evaluation lacks a standard protocol, 
we adopt dataset-specific procedures based on the available annotations. 
\textbf{speechocean762:} ASR and reference words are Levenshtein-aligned, 
and PCC is reported over matched+substituted (M+S), matched (M), and substituted (S) conditions. M+S covers 98.37\%, 98.60\%, and 98.95\% of the reference words 
for medium.en, large-v3, and Qwen3-ASR-0.6B~\cite{shi2026qwen3asr} transcripts, respectively.
\textbf{MultiPA:} Since no reference transcript is available, we follow its original protocol~\cite{chen2024multipa}, 
assigning each ASR word the duration-weighted mean of overlapping time-aligned human
ratings before computing PCC.

% TODO: Verify the script used to report separate M-only and S-only PCCs.
% openpa-main currently exposes Levenshtein reference-to-ASR target mapping,
% but the inspected evaluation path does not separately report these subsets.

\subsection{Evaluation results}

% TODO: Add quantitative results and analysis.
\noindent\textbf{In-domain:} As shown in Table~\ref{tab:speechocean-results}, OpenPA 
leads in overall averaged PCC for both closed and simulated open responses, 
driven by significant word-level improvements. With ground-truth transcripts, 
fluency and prosody slightly lag behind HiPPO$^{*}$, potentially due to the lack of dedicated suprasegmental features. 
Yet OpenPA scores both levels without forced alignment or separate feature extraction 
and retains higher word PCC, especially under ASR substitutions.

\noindent\textbf{Ablation:} Table~\ref{tab:ablation} compares different training strategies.
GT-only training yields the highest matched-word (M) PCC, but its low substitution (S) PCC of 0.076-0.152 
shows little ability to score ASR-substituted words. Mixed-transcript adaptation improves M+S, S, and averaged utterance-level PCC, 
with adaptive weighting providing further S gains. Notably, OpenPA benefits from higher-quality transcripts produced by advanced ASR systems, 
even when the recognizers are unseen during training: the lowest-WER Qwen3-ASR-0.6B yields the highest M+S and averaged utterance-level PCC.

\noindent\textbf{Cross-dataset:} Beyond the forced-alignment-based MultiPA$^{*}$, 
we compare OpenPA with TextPA~\cite{chen2025read}, 
a recent zero-shot LLM-based approach leveraging textual speech descriptions. 
As shown in Table~\ref{tab:multipa-results}, OpenPA outperforms MultiPA$^{*}$ at both granularities 
and both TextPA variants on shared utterance-level metrics. Notably, prosodic PCC improves from 0.473 (MultiPA$^{*}$) to 0.644, 
although the relatively lower word- and prosody-level correlations may reflect the domain shift from read-aloud training speech 
to conversational speech collected from L2 learners interacting with an English-learning chatbot.

\noindent\textbf{Cross-task:} To our knowledge, direct transfer from L2 pronunciation assessment to pathological speech remains underexplored. 
Table~\ref{tab:pathbench-results} compares OpenPA with reference-free ASRIC, DArtP, and ASR confidence, alongside reference-text PER and ArtP~\cite{halpern2026pathbench}. 
Without pathological-speech training, OpenPA achieves the highest speaker-level PCC of 0.82 using averaged fluency predictions. 
As oral-cancer severity ratings consider speaking rate~\cite{halpern2023oral}, shared temporal cues may explain this transfer, 
yet clinical validity remains unestablished.

\begin{table}[!t]
  \centering
  \setlength{\textfloatsep}{0pt}
  \setlength{\belowcaptionskip}{3pt}
  \vspace{-7pt}
  \caption{In-domain ablation PCC. Seen/Unseen refers to mixed-adaptation ASR sources;
  Full adds adaptive substitution weighting. Utt. Avg. averages four utterance PCCs.}
  \label{tab:ablation}
  \footnotesize
  \setlength{\tabcolsep}{0.5pt}
  \renewcommand{\arraystretch}{0.85}
  \begin{tabular*}{\columnwidth}{@{\extracolsep{\fill}}l@{\hspace{1pt}}ll@{\hspace{1pt}}cc@{}}
    \toprule
    Transcript & Type & Training & Word Acc. (M+S/M/S)↑ & Utt. Avg.↑ \\
    \midrule
    medium.en & & GT-only & 0.416\,/\,\textbf{0.479}\,/\,0.118 & 0.765 \\
    \scriptsize(WER: 20.1\%$\downarrow$) & Seen & Mixed & 0.504\,/\,0.471\,/\,0.287 & 0.779 \\
              & & Full & \textbf{0.506}\,/\,0.451\,/\,\textbf{0.295} & \textbf{0.779} \\
    \midrule
    large-v3 & & GT-only & 0.458\,/\,\textbf{0.546}\,/\,0.152 & 0.764 \\
    \scriptsize(WER: 16.3\%$\downarrow$) & Unseen & Mixed & 0.526\,/\,0.521\,/\,0.328 & 0.777 \\
             & & Full & \textbf{0.526}\,/\,0.504\,/\,\textbf{0.335} & \textbf{0.780} \\
    \midrule
    Qwen3-ASR-0.6B & & GT-only & 0.452\,/\,\textbf{0.547}\,/\,0.076 & 0.779 \\
    \scriptsize(WER: 12.4\%$\downarrow$) & Unseen & Mixed & 0.540\,/\,0.534\,/\,0.299 & 0.786 \\
                   & & Full & \textbf{0.544}\,/\,0.512\,/\,\textbf{0.333} & \textbf{0.787} \\
    \bottomrule
  \end{tabular*}
  \vspace{-8pt}
\end{table}

\vspace{0pt}
\subsection{Qualitative analysis}
\footnotetext{\href{https://fuann.github.io/open-apa/audios/}{Listen to the audio sample.}}

Figure~\ref{fig:qualitative-attention} compares cross-attention patterns under reference and large-v3 transcripts, 
where the latter substitutes \emph{person} with \emph{pizza}. Low scores for \emph{always} (4.29) and \emph{thought} (3.95) 
coincide with diffuse or shifted attention, unlike the localized peaks of well-pronounced words. 
Under the ASR transcript, earlier attention patterns and scores remain stable, while the final score changes only from 4.20 to 4.64. 
These patterns illustrate how score tokens can attend to relevant acoustic regions without explicit word boundaries, even under ASR substitutions.

\vspace{-10pt}
\section{Conclusion}
\label{sec:conclusion}

We presented OpenPA, a unified framework for word- and utterance-level pronunciation assessment without forced alignment 
across closed- and open-response settings. Through mixed-transcript adaptation, OpenPA improves robustness to ASR-generated transcripts, 
including those from unseen recognizers. Moreover, in-domain and zero-shot MultiPA results demonstrate strong assessment performance and cross-dataset generalisation. 
Exploratory oral-cancer evaluation further suggests cross-task potential. 
Future work will investigate larger atypical-speech cohorts and finer phonetic granularity.

\newpage
% References should be produced using the bibtex program from suitable
% BiBTeX files (here: strings, refs, manuals). The IEEEbib.bst bibliography
% style file from IEEE produces unsorted bibliography list.
% -------------------------------------------------------------------------
\newcounter{openpabibitem}
\AtBeginEnvironment{thebibliography}{%
  \renewcommand{\baselinestretch}{1.15}\selectfont
  \setcounter{openpabibitem}{0}%
  \let\openpaorigbibitem\bibitem
  \renewcommand{\bibitem}[1]{%
    \stepcounter{openpabibitem}%
    \def\openpabibbreak{}%
    \ifnum\value{openpabibitem}=1
      \setlength{\itemsep}{10.4pt}%
    \fi
    \ifnum\value{openpabibitem}=9
      \def\openpabibbreak{\newpage\setlength{\itemsep}{4pt}}%
    \fi
    \openpabibbreak
    \openpaorigbibitem{#1}%
  }%
}
\bibliographystyle{IEEEbib}
\bibliography{strings,refs}

\end{document}